\documentclass[prb]{revtex4}

\usepackage{amsmath,amssymb}

\usepackage{graphicx}

\begin{document}

\widetext

\title{Exciton Transfer Integrals Between Polymer Chains}

\author{William Barford\footnote{E.mail address: william.barford@chem.ox.ac.uk}}

\affiliation{Physical and Theoretical Chemistry Laboratory,
University of Oxford, South Parks  Road, Oxford, OX1 3QZ, United
Kingdom}

\begin{abstract}
The line-dipole approximation for the evaluation of the exciton
transfer integral, $J$, between conjugated polymer chains is
rigorously  justified. Using this approximation, as well as the
plane-wave approximation for the exciton center-of-mass
wavefunction, it is shown analytically that $J \sim L$ when the
chain lengths are smaller than the separation between them, or $J
\sim L^{-1}$  when the chain lengths are larger than their
separation, where $L$ is the polymer length. Scaling relations are
also obtained numerically for the more realistic standing-wave
approximation for the exciton center-of-mass wavefunction, where it
is found that for chain lengths larger than their separation $J \sim
L^{-1.8}$ or $J \sim L^{-2}$, for parallel or collinear chains,
respectively. These results
 have important implications for the
photo-physics of conjugated polymers and self-assembled molecular
systems, as the Davydov splitting in aggregates and the F\"orster
transfer rate for exciton migration decreases with chain lengths
larger than their separation. This latter result has obvious
deleterious consequences for the performance of polymer photovoltaic
devices.
\end{abstract}


\maketitle

\section{Introduction}

The transport of excitons - and hence energy - through molecular
materials is determined by the exciton transfer integral between
neighboring conjugated chromophores. Typically, the length of these
chromophores are larger than the distances between them, and thus
the usual expression for the transfer integrals, derived by invoking
the dipole approximation, is invalid. Indeed, the use of the dipole
approximation implies that the exciton transfer integral between
equivalent neighboring chromophores scales linearly as the
chromophore length. This prediction contradicts the computations of
the transfer integrals using various computational techniques, which
predict that the transfer integrals vanish in the asymptotic
limit\cite{soos90, spano97, spano98, cornil98, beljonne00} - a
result in agreement with the decreasing Davydov splitting as a
function of chromophore length observed in self-assembled molecular
aggregates\cite{westenhoff06}.

The exciton transfer integral, $J_{mn}$, between two polymer
chains (or linear conjugated chromophores) $m$ and $n$ is defined
by\cite{book},
\begin{eqnarray}\label{Eq:1}
    J_{mn} = \sum_{i\in m j\in n}V_{ij} \left[
    {_m}\langle \textrm{GS} |\hat{N}_i| \textrm{EX} \rangle_m\right]
    \left[{_n}\langle \textrm{EX} |\hat{N}_j|\textrm{GS}\rangle_n
    \right],
\end{eqnarray}
where $|\textrm{GS}\rangle_m$ and $|\textrm{EX}\rangle_m$ are the
ground and excited states of polymer $m$.  $V_{ij}$ is the Coulomb
interaction between electrons in orbitals $\phi_i$ and $\phi_j$,
and $\hat{N}_i$ is the number operator for electrons in orbital
$\phi_i$. ${_m}\langle \textrm{GS} |\hat{N}_i| \textrm{EX}
\rangle_m$ is thus the transition density for orbital $i$ on
polymer $m$. When $|\textbf{r}_i -\textbf{r}_j|$ is large compared
to the interatomic spacing the Coulomb potential is
\begin{equation}\label{Eq:2}
    V_{ij}=
    \frac{\textrm{e}^2}{|\textbf{r}_i -
    \textbf{r}_j|},
\end{equation}
where $\textbf{r}_i$ is the coordinate of atom $i$.

In practice, the exact computation of exciton transfer integrals
using transition densities is computationally expensive. A
convenient approximation is the so-called line-dipole
approximation\cite{grage03, beenken04}, which provides a
physically intuitive, yet accurate description for exciton
transfer when the chromophore separation is large enough
(typically, three or four times the monomer length).

In this paper we use the exciton model to explicitly derive the
line-dipole approximation. Next, using the line-dipole
approximation, analytical expressions for the exciton transfer
integral between parallel and collinear conjugated polymers are
derived, and the scaling with polymer length is determined. We
conclude by discussing the implications of these results for exciton
transfer via the F\"orster mechanism.

Before deriving the line-dipole approximation, however, we review
the point-dipole approximation for the evaluation of $J_{mn}$. We
define $\tilde{\textbf{r}}_i$ and $\tilde{\textbf{r}}_j$ as the
site coordinates relative to the center-of-mass of their
respective molecules,
\begin{eqnarray}
\tilde{\textbf{r}}_i = \textbf{r}_i - \textbf{R}_m, \nonumber
\\
 \tilde{\textbf{r}}_j = \textbf{r}_j - \textbf{R}_n,
\end{eqnarray}
where $\textbf{R}_m$ and $\textbf{R}_n$ are the center-of-mass
coordinates of molecules $m$ and $n$, respectively. Then, if
\begin{equation}\label{Eq:3}
|\tilde{\textbf{r}}_i - \tilde{\textbf{r}}_j| << |\textbf{R}_m -
\textbf{R}_n| \equiv |\textbf{R}_{mn}|
\end{equation}
we may perform the dipole approximation and write,
\begin{equation}\label{}
    \sum_{i\in m
    j\in n}
    \frac{1}{|\textbf{r}_{i} -
    \textbf{r}_{j}|} \approx
    \frac{\sum_{i\in
    m}\tilde{\textbf{r}}_i\cdot\sum_{j\in
    n}\tilde{\textbf{r}}_j}{|\textbf{R}_{mn}|^3} -
    \frac{3\left(\sum_{i\in m}
    \textbf{R}_{mn}\cdot\tilde{\textbf{r}}_i\right)
    \left(\sum_{j\in n}
    \textbf{R}_{mn}\cdot\tilde{\textbf{r}}_j\right)}
    {|\textbf{R}_{mn}|^5}.
\end{equation}
Finally, substituting into Eq.\ (\ref{Eq:1}) gives,
\begin{equation}\label{Eq:4}
    J_{mn}= \kappa_{mn}J^0_{mn},
\end{equation}
where
\begin{equation}\label{Eq:5}
   J^0_{mn}=
 \frac{\left[{_m}\langle \textrm{GS} |\hat{\mu}_m| \textrm{EX} \rangle_m\right]\left[{_n}\langle
 \textrm{EX}|\hat{\mu}_n|\textrm{GS}
    \rangle_n\right]}
    {|\textbf{R}_{mn}|^3},
\end{equation}
and
\begin{equation}\label{Eq:6}
     \kappa_{mn} =\hat{\textbf{r}}_m\cdot \hat{\textbf{r}}_n - 3
     (\hat{\textbf{R}}_{mn}\cdot\hat{\textbf{r}}_m)
     (\hat{\textbf{R}}_{mn}\cdot\hat{\textbf{r}}_n),
\end{equation}
is an orientational factor. $\hat{\textbf{r}}_m$ and
$\hat{\textbf{R}}_{mn}$ are the unit vector parallels to
$\hat{\mu}_m$ and $\textbf{R}_{mn}$, respectively. $ \hat{\mu}_m$ is
the electronic dipole operator for molecule $m$, defined by
\begin{equation}\label{Eq:7}
   \hat{\mu}_m = \textrm{e}\sum_{i\in
    m}\tilde{\textbf{r}}_i\hat{N}_i,
\end{equation}
and
\begin{equation}\label{Eq:10}
   _m\langle \textrm{GS} |\hat{\mu}_m| \textrm{EX} \rangle_m \equiv
   \mu_N
\end{equation}
is the transition dipole moment of polymer $m$ with $N$ repeat
units.

In practice the condition Eq.\ (\ref{Eq:3}) is far too severe for
conjugated polymers in the solid state, and Eq.\ (\ref{Eq:5}) is
not applicable. However, the much weaker condition that,
\begin{equation}\label{Eq:8}
d <<  |\textbf{R}_{mn}|,
\end{equation}
where $d$ is the monomer size is often satisfied, and under this
condition the line-dipole approximation becomes valid.

\section{The Line Dipole Approximation}\label{Se:2}

\begin{figure}[tb]
\begin{center}
\includegraphics[scale=1]{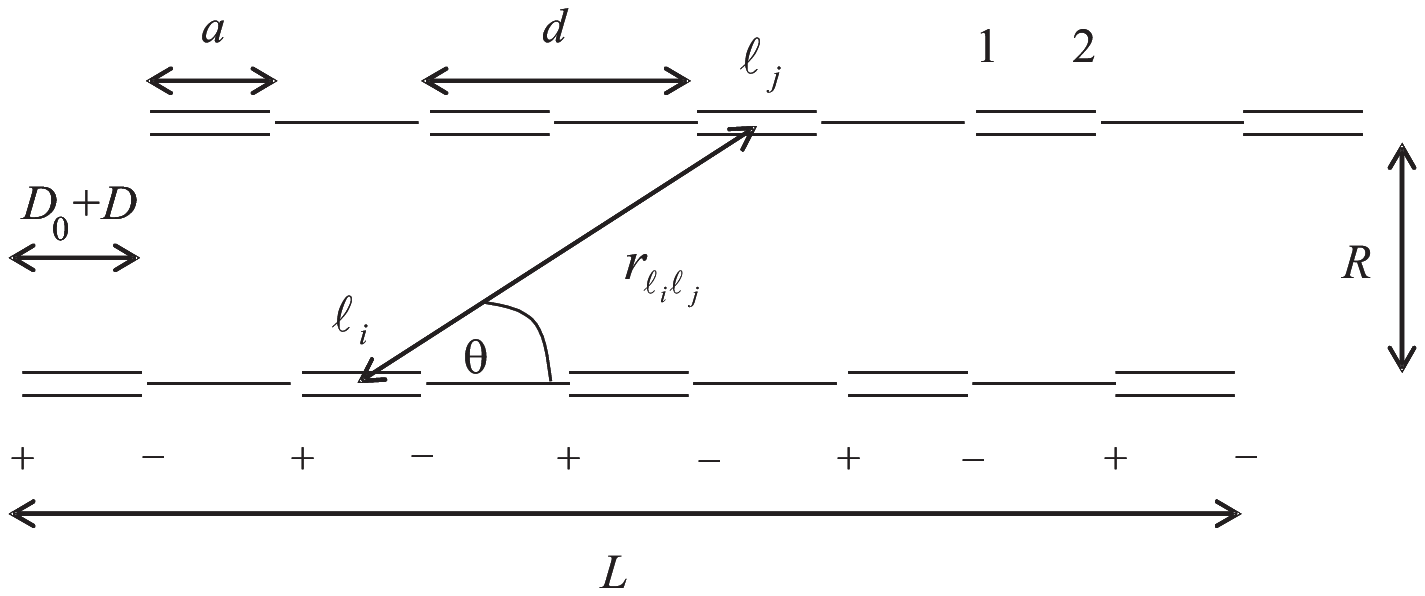}
\end{center}
\caption{} \label{Fi:1}
\end{figure}

In this section the line-dipole approximation will be justified for
a simplified model of polymers, namely a chain of dimers (or double
bonds) connected by single bonds, as illustrated in Fig.\
\ref{Fi:1}. This model is of course applicable to polyacetylene.
However, more generally it is also applicable if the dimer
represents a monomer.

To derive the line-dipole approximation it is also necessary to
introduce a model for excitons in conjugated polymers. In the
weak-coupling limit, defined by the electronic band width being
greater than the Coulomb interaction, the excited states of
conjugated polymers are Mott-Wannier excitons\cite{book, barford02},
defined by
\begin{equation}\label{Eq:12}
    |EX\rangle = \sum_{r,R} \psi(r)\Psi(R)|R,r \rangle,
\end{equation}
where $ |R,r \rangle$ is an electron-hole basis state,
\begin{equation}
  |R,r \rangle = \frac{1}{\sqrt{2}} \left( c_{R+r/2, \uparrow}^{c \dagger} c_{R-r/2, \uparrow}^v
  + c_{R+r/2, \downarrow}^{c \dagger} c_{R-r/2, \downarrow}^v \right)|\textrm{GS}\rangle.
\end{equation}
$c_{R+r/2, \sigma}^{c \dagger}$ creates an electron with spin
$\sigma$ in the conduction band Wannier orbital at $R+r/2$, while
$c_{R-r/2, \sigma}^v$ destroys an electron with spin $\sigma$ in the
valence band Wannier orbital at $R-r/2$, and $|\textrm{GS}\rangle$
is the ground state. $\psi(r)$ is the `hydrogenic' wavefunction for
the particle-hole pair, where $r$ is the relative coordinate.
$\Psi(R)$ is the center-of-mass envelope wavefunction, where $R$ is
the center-of-mass coordinate.

To a good approximation it can be shown\cite{book} that the
transition densities satisfy,
\begin{equation}\label{Eq:17a}
   \langle \textrm{GS} |\hat{N}_i| \textrm{EX} \rangle =
   \Psi(R_{\ell_i})\left(\frac{\psi(0)}{\sqrt{2}}\right)(-1)^{i+1}.
\end{equation}
We note that the transition densities are modulated by the
center-of-mass wavefunction and alternate in sign, as illustrated
for the lowest excited exciton in Fig.\ \ref{Fi:1}.

Using Eq.\ (\ref{Eq:17a}), Eq.\ (\ref{Eq:1}) now becomes,
\begin{eqnarray}\label{Eq:}
    J_{mn} = \left(\frac{\psi(0)^2}{2}\right)\sum_{i\in m} \sum_{j\in n} V_{ij}\Psi(R_{\ell_i})
    \Psi(R_{\ell_j})(-1)^{i+1}(-1)^{j+1},
\end{eqnarray}
To simplify this expression it is convenient to partition the sum
over sites, $i$, as a sum over unit cells, $\ell$, and a sum over
sites within a unit cell:
\begin{eqnarray}\label{Eq:}
    J_{mn} = \left(\frac{\psi(0)^2}{2}\right)\sum_{\ell_i\in m} \sum_{\ell_j\in n}\Psi(R_{\ell_i})
    \Psi(R_{\ell_j})\left\{\sum_{i=1,2}
    \sum_{j=1,2}V_{ij}(-1)^{i+1}(-1)^{j+1}\right\}.
\end{eqnarray}
Then, if the unit cell size, $d$, satisfies,
\begin{equation}\label{Eq:}
d <<  r_{\ell_i \ell_j},
\end{equation}
where $r_{\ell_i \ell_j}$  is the distance between the unit cells,
the term in curly parentheses can now be simplified using the
dipole approximation to become,
\begin{equation}\label{Eq:}
\sum_{i=1,2} \sum_{j=1,2}V_{ij}(-1)^{i+1}(-1)^{j+1}=
\frac{\textrm{e}^2 a^2 \kappa_{\ell_i \ell_j}}{r^3_{\ell_i \ell_j}},
\end{equation}
and thus,
\begin{eqnarray}\label{Eq:22}
    J_{mn} = \mu_1^2\sum_{\ell_i\in m} \sum_{\ell_j\in n}\Psi(R_{\ell_i})
    \frac{\kappa_{\ell_i \ell_j}}{r^3_{\ell_i
    \ell_j}}\Psi(R_{\ell_j}).
\end{eqnarray}
$\mu_1$ is the transition dipole moment of a dimer, defined by
\begin{equation}\label{Eq:}
\mu_1 = \frac{\textrm{e} a \psi(0) }{\sqrt{2}}
\end{equation}
and $a$ is the bond length. Since $\mu_1$ is independent of the
length of the polymer, Eq.\ (\ref{Eq:22}) reveals how the exciton
transfer integral varies with the polymer length, as will be
described in the next section. However, an alternative
representation of Eq.\ (\ref{Eq:22}) will now be derived which makes
a more direct comparison to the point-dipole approximation and
introduces the line-dipole approximation.

Using the expression for the transition dipole moment of the whole
chain (from Eq.\ (\ref{Eq:7}),  Eq.\ (\ref{Eq:10}), and Eq.\
(\ref{Eq:17a}))\cite{book}, namely
\begin{equation}\label{Eq:24}
\mu_N = \mu_1{\sum_{\ell}\Psi(R_{\ell})},
\end{equation}
where the sum is over unit cells, Eq.\ (\ref{Eq:22}) can be
re-expressed as:
\begin{eqnarray}\label{Eq:25}
    J_{mn} = \left(\frac{\mu_N}{\sum_{\ell}\Psi(R_{\ell})}\right)^2\sum_{\ell_i\in m} \sum_{\ell_j\in n}\Psi(R_{\ell_i})
    \frac{\kappa_{\ell_i \ell_j}}{r^3_{\ell_i
    \ell_j}}\Psi(R_{\ell_j}).
\end{eqnarray}
This expression for the transfer integral between conjugated
polymers justifies the line-dipole approximation as may be seen as
follows. Using Eq.\ (\ref{Eq:24}), the transition dipole moment for
a polymer of $N$ monomer units may be written as a sum of the
transition dipole moments for each monomer, namely
\begin{equation}\label{Eq:}
\mu_N = {\sum_{\ell}\mu(R_{\ell})},
\end{equation}
where
\begin{equation}\label{Eq:27}
\mu(R_{\ell}) \equiv \mu_1 \Psi(R_{\ell}) = \mu_N
\frac{\Psi(R_{\ell})}{\sum_{{\ell}}\Psi(R_{\ell})}.
\end{equation}
Assuming that each monomer of polymer $m$ interacts with each
monomer of polymer $n$ via the point-dipole approximation, the total
exciton integral between the polymers is\cite{beenken04},
\begin{eqnarray}\label{Eq:28}
    J_{mn} = \sum_{\ell_i\in m} \sum_{\ell_j\in n}\mu(R_{\ell_i})
    \frac{\kappa_{\ell_i \ell_j}}{r^3_{\ell_i
    \ell_j}}\mu(R_{\ell_j}).
\end{eqnarray}
Eq.\ (\ref{Eq:25}) is thus derived by inserting Eq.\ (\ref{Eq:27})
into Eq.\ (\ref{Eq:28}).

Before discussing the evaluation of Eq.\ (\ref{Eq:22}) and its
validity, we describe the standing-wave and plane-wave
approximations for the center-of-mass wavefunction,
$\Psi(R_{\ell})$.

\subsection{Standing waves}

For linear polymers with $N$ equivalent monomers and open boundary
conditions the exciton center-of-mass wavefunction satisfies,
\begin{equation}\label{Eq:29}
\Psi(R_{\ell}) = \sqrt{\frac{2}{N+1}}\sin\left(\beta
R_{\ell}\right),
\end{equation}
where $\beta$ is the pseudo-momentum of the center-of-mass,
defined by $\beta = \pi j\ /(N+1)$ with $ 1 \le j \le N$. This
form of $\Psi(R_{\ell})$ in Eq.\ (\ref{Eq:24}) gives
\begin{eqnarray}\label{Eq:}
\mu_N = && \mu_1 \sqrt{\frac{2}{N+1}}\cot(\beta/2)
\\
\nonumber \sim && \sqrt{N}/j,
\end{eqnarray}
when $N/j \gg 1$. We therefore see that in the point-dipole
approximation $J \sim L$.

\subsection{Plane waves}

In the next section it will be expedient to employ the plane wave
approximation for $\Psi(R_{\ell})$ valid for translationally
invariant systems,
\begin{equation}\label{Eq:31}
\Psi(R_{\ell}) = \frac{\exp(iKR_{\ell})}{{\sqrt{N}}},
\end{equation}
where $K$ is the Bloch momentum of the center-of-mass, defined by $K
= 2 \pi j/N d$ with $-N/2 \le j \le N/2$. This approximation allows
analytical expressions to be derived for the exciton transfer
integral, although it is less applicable than the standing wave
approximation in linear conjugated polymers and chromophores.

\subsection{Validity of the Line-dipole approximation}

The validity of the line-dipole approximation  has already been
discussed elsewhere (see ref.\cite{grage03} and references therein).
 When $R = d$ there is a
significant error in the line-dipole approximation result. However,
for $R \gtrsim 2d$ the line-dipole approximation result is in good
agreement with the exact result for short chains, becoming
essentially exact for all chain lengths when $R \gtrsim 3d$.

\subsection{Parallel Chains}

An analytical expression can be derived for the transfer integral in
the continuum limit assuming the plane wave form for the
center-of-mass wavefunctions, Eq.\ (\ref{Eq:31}). Then,
\begin{equation}\label{Eq:33}
J \approx  \frac{\mu_1^2}{Ld}\int_0^L  \int_0^L
\frac{(1-3\cos^2\theta)}{r^3}dx_1 dx_2,
\end{equation}
where $r$ is the distance between the segments at $x_1$ on chain 1
and $x_2$ on chain 2. For parallel chains with $D=D_0=0$ the
computation of Eq.\ (\ref{Eq:33}) gives,
\begin{equation}\label{Eq:}
J =  \left(\frac{2\mu_1^2}{R L d}\right)
\left(1-\frac{R}{(R^2+L^2)^{1/2}}\right).
\end{equation}

The dependence of the exciton transfer integral on the polymer
length is made more explicit by defining $J$ as,
\begin{equation}\label{Eq:}
J =  \left(\frac{2\mu_1^2}{R^2 d}\right) g(\tilde{L}),
\end{equation}
where $g(\tilde{L})$, the dimensionless exciton transfer integral,
is
\begin{equation}\label{Eq:}
g(\tilde{L}) = \frac{1}{\tilde{L}}\left( 1-
\frac{1}{\sqrt{(1+\tilde{L}^2)}}\right)
\end{equation}
and $\tilde{L} = L/R$.

$g(\tilde{L})$ is plotted versus $\tilde{L}$ for $R=5d$ in Fig.
\ref{Fi:3} and Fig. \ref{Fi:4}. As $\tilde{L} \rightarrow 0$ (or
$L << R$) $g(\tilde{L}) \rightarrow \tilde{L}/2$ and thus $J \sim
LR^{-3}$, confirming the point-dipole approximation for the whole
chain. Conversely, as $\tilde{L} \rightarrow \infty$ (or $L >> R$)
$g(\tilde{L}) \rightarrow \tilde{L}^{-1}$ and thus $J \sim
(RL)^{-1}$. The maximum value of $g$ is $g_{\textrm{max}} \approx
0.300$, occurring at
\begin{equation}\label{Eq:}
\tilde{L}_{\textrm{max}} = \left(\frac{1 +
\sqrt{5}}{2}\right)^{1/2} \approx 1.2720\cdots
\end{equation}
These idealized analytical results are consistent with the quantum
chemistry calculations on explicit polymer systems of other
authors\cite{spano97, spano98, cornil98, beljonne00}.

It is also instructive to express $J$ as,
\begin{equation}\label{Eq:}
J = \left(\frac{\mu_N^2}{R^3}\right) f(\tilde{L}),
\end{equation}
where
\begin{equation}\label{Eq:}
f(\tilde{L}) = \frac{2}{\tilde{L}^2}\left( 1-
\frac{1}{\sqrt{(1+\tilde{L}^2)}}\right)
\end{equation}
can be regarded as a correction to the point-dipole approximation
for the whole chain. As $\tilde{L} \rightarrow 0$ $f(\tilde{L})
\rightarrow 1$, while
 as  $\tilde{L} \rightarrow \infty$ $f(\tilde{L}) \rightarrow
2\tilde{L}^{-2}$. $f(\tilde{L})$ is plotted in Fig.\ \ref{Fi:5}.

The analytical results derived in this section have assumed the
plane-wave approximation for the exciton center-of-mass
wavefunction. A more realistic approximation for linear polymer
chains, however, is the standing-wave approximation, Eq.\
(\ref{Eq:29}). Analytical expressions cannot be derived using this
form of $\Psi$. However, Eq. (\ref{Eq:22}) is easily evaluated
numerically. Fig. \ref{Fi:3} and Fig.\ \ref{Fi:4} show
$g(\tilde{L})$ in this approximation. For $\tilde{L} \lesssim 20$
the plane and standing wave approximations agree rather well.
However, the asymptotic behavior is different, decreasing more
rapidly with chain length for the standing wave approximation,
namely $g(\tilde{L}) \sim \tilde{L}^{-1.8}$, as indicated in Fig.\
\ref{Fi:4}.

\subsection{Collinear Chains}

For collinear chains with $R=0$, $D_0=L$, and $D$ arbitrary, the
evaluation of Eq.\ (\ref{Eq:33}) for the plane-wave approximation
gives,
\begin{equation}\label{Eq:}
J =  -\left(\frac{2\mu_1^2}{D^2 d}\right) \frac{L}{(L+D)(2L+D)},
\end{equation}
or
\begin{equation}\label{Eq:}
J =  -\left(\frac{2\mu_1^2}{D^2 d}\right) g(\tilde{L})
\end{equation}
where
\begin{equation}\label{Eq:}
g(\tilde{L}) = \frac{\tilde{L}}{(\tilde{L}+1)(2\tilde{L}+1)}
\end{equation}
and $\tilde{L} =L/D$. $g(\tilde{L})$ is plotted versus $L/D$ for
$D=5d$ in Fig. \ref{Fi:6} and Fig. \ref{Fi:7}. As $\tilde{L}
\rightarrow 0$ $g(\tilde{L}) \rightarrow \tilde{L}$, while as
$\tilde{L} \rightarrow \infty$ $g(\tilde{L}) \rightarrow
(2\tilde{L})^{-1}$. The maximum value of $g$ is $g_{\textrm{max}}
\approx 0.172$, occurring at $\tilde{L}_{\textrm{max}} =
1/\sqrt{2}$.

Fig. \ref{Fi:6} and Fig.\ \ref{Fi:7} also show the numerical
evaluation of the transfer integral using the standing-wave
approximation for the exciton center-of-mass wavefunction. In this
case, as $\tilde{L} \rightarrow \infty$ $g(\tilde{L}) \sim
\tilde{L}^{-2}$.

For collinear chains $J$ can also be expressed as a correction to
the point-dipole approximation for the whole chain as,
\begin{equation}\label{Eq:}
J = -\left(\frac{2\mu_N^2}{D^3}\right) f(\tilde{L}),
\end{equation}
where
\begin{equation}\label{Eq:}
f(\tilde{L}) = \frac{1}{(\tilde{L} +1)(2\tilde{L}+1)}.
\end{equation}
This function is plotted in Fig. \ref{Fi:5}.

\section{Discussion and Concluding Remarks}

In this paper the line-dipole approximation for the evaluation of
the exciton transfer integral, $J$, between conjugated polymer
chains has been rigorously justified for a chain of dimers. Quite
generally, it is physically reasonable to apply this approximation
to a conjugated polymer or linear chromophore with general
monomeric units.

Then, using the line-dipole approximation, as well as the plane-wave
approximation for the exciton center-of-mass wavefunction, it has
been shown analytically that $J \sim L$ when the chain lengths are
smaller than the separation between them, where $L$ is the polymer
length. However, when the chain lengths are larger than their
separation, $J$ is a decreasing function of chain length, scaling as
$L^{-1}$ for the plane-wave approximation. Scaling relations have
also been obtained numerically for the more realistic standing-wave
approximation for the exciton center-of-mass wavefunction, where it
is found that for chain lengths larger than their separation $J \sim
L^{-1.8}$ or $J \sim L^{-2}$, for parallel or collinear chains,
respectively.

The scaling of $J \sim L$ for short chain lengths is a reflection of
the fact that at these lengths scales the point-dipole approximation
may be applied to the entire chain, implying that $J \propto \langle
\textrm{GS} |\hat{\mu}| \textrm{EX} \rangle^2 \sim L$. Similarly,
the scaling of $J \sim L^{-1}$ for collinear chains in the
plane-wave approximation is easy to understand for chain lengths
large compared to their separation. In this case the exciton dipoles
are uniformally distributed along both chains of length $L$. The
double line integral of $r^{-3}$ trivially gives the $L^{-1}$
scaling.

The scaling of $J$ with $L$ for long \textit{parallel} chains is
rather counter-intuitive, however, as it implies (as discussed
below) that the probability of exciton transfer between neighboring
parallel chains is a decreasing function of length. The physical
origin of this behavior can be understood in two ways. First,
consider the exciton transfer integral for parallel chains when
their separation is of the order of the monomer length ($d$). In
this limit $J$ is a periodic function of the relative shift variable
$D$\cite{soos90}. Fig.\ \ref{Fi:8} shows $J$ for $D = 0$ (the
in-phase configuration), $D=a$ (the out of phase configuration), and
$D=a/2$ (the phase-cancelation configuration), illustrating this
periodic variation. As the separation between parallel chains
increases this periodic variation vanishes because of interference
effects from longer distance transition densities. It is these
interference effects that mean ultimately the exciton transfer
integral must vanish in the asymptotic limit for chain separations
greater than a few monomer lengths.

A second way to motivate this scaling behavior is to consider the
sign of $J$ as two finite-length dipoles are moved longitudinally
relative to another. As they slide passed each other the sign of $J$
will change once the magic angle ($\theta = \cos^{-1}(1/\sqrt{3})$)
is reached. Evidently, in the limit of infinitely long chains this
thought-experiment is nonsensical, implying that $J=0$.

The results for the exciton transfer integral have been derived
using the assumption that the exciton center-of-mass wavefunction,
$\Psi$, is described either by a standing wave or a plane wave. The
standing wave approximation is more relevant than the plane wave
approximation in linear conjugated polymers or chromophores,
although the plane wave approximation enables analytical results to
be derived. In reality, however, disorder or self-trapping will
modify the exciton center-of-mass wavefunction from both of these
idealized functional forms, and under these circumstances it is
necessary to evaluate Eq.\ (\ref{Eq:22}) with the explicit form of
$\Psi$. Such a calculation is beyond the scope of the present paper.

The results presented in this paper have important implications for
the photo-physics of conjugated polymers and self-assembled
molecular systems. First, as already reported\cite{soos90, spano97,
spano98, cornil98, beljonne00}, the Davydov splitting in aggregates
is a decreasing function of chain length for chain lengths larger
than their separation. Second, the F\"orster transfer rate, $k$, is
reduced by a factor of $f^2(\tilde{L})$, where $f(\tilde{L})$ is the
correction to the point-dipole approximation for the entire chain,
shown in Fig.\ \ref{Fi:5}. This conclusion may be understood by
noting that in the point-dipole approximation,
\begin{equation}\label{Eq:}
k \propto \frac{1}{R^6} \int_0^{\infty} \frac{I_D(\omega)
\alpha_A(\omega)}{\omega^4}d\omega,
\end{equation}
where $I_D(\omega)$, the donor emission spectrum, and
$\alpha_A(\omega)$, the acceptor absorption spectrum, are both
proportional to the square of the transition dipole moments. In the
line-dipole approximation, therefore, there is a correction of
$f(\tilde{L})$ for each of these squared transition dipole moments.
This latter result has obvious important deleterious consequences
for the performance of polymer photovoltaic devices where exciton
migration occurs via F\"orster transfer between neighboring
conjugated polymers or chromophores.

\begin{acknowledgements}
The author acknowledges the financial support of the EPSRC (grant
ref EP/D038553/1).
\end{acknowledgements}

\pagebreak
\begin{center}
\bf
{Figure Captions}
\end{center}

\begin{figure}[tb]
\begin{center}
\includegraphics[scale=1]{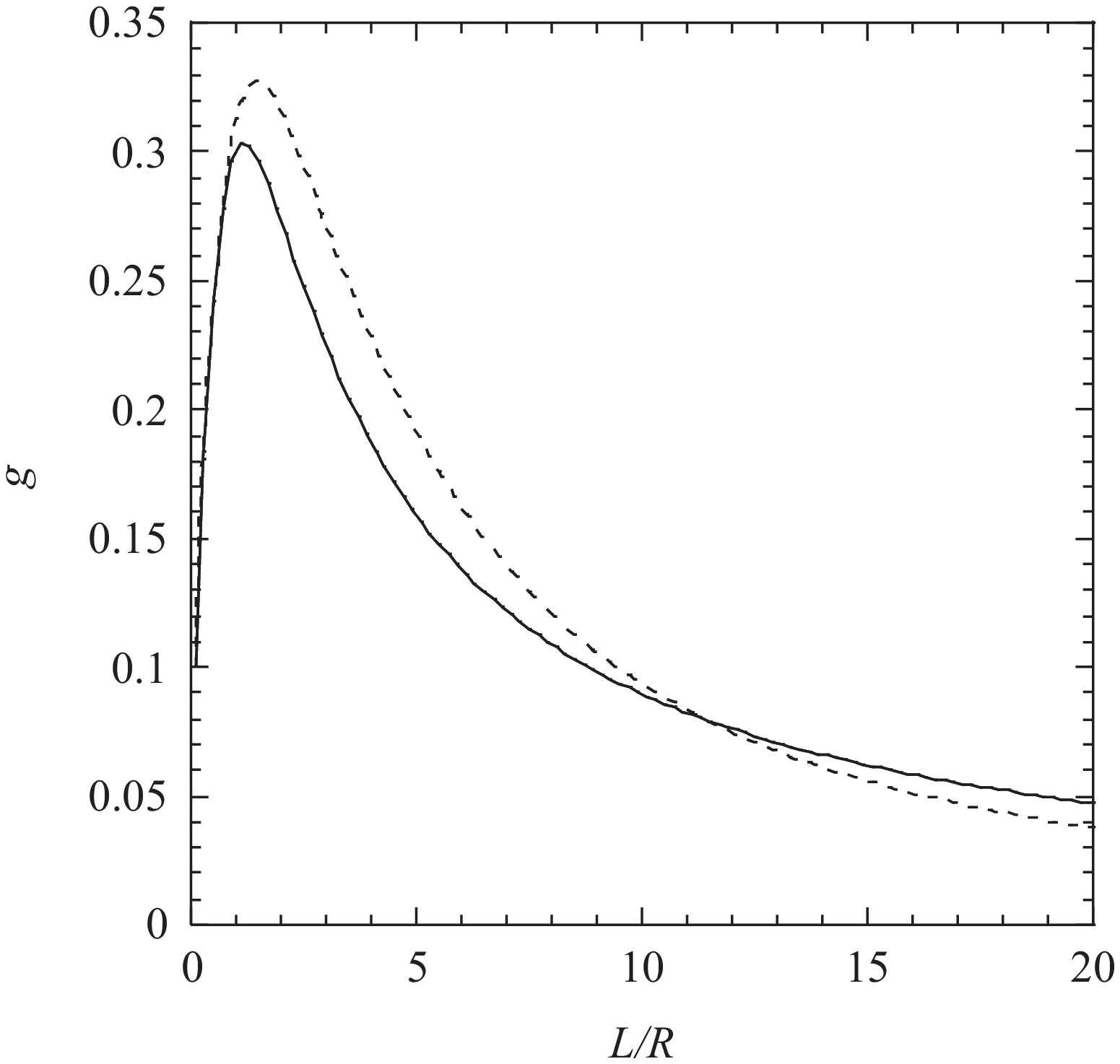}
\end{center}
\caption{} \label{Fi:3}
\end{figure}

\begin{figure}[tb]
\begin{center}
\includegraphics[scale=1]{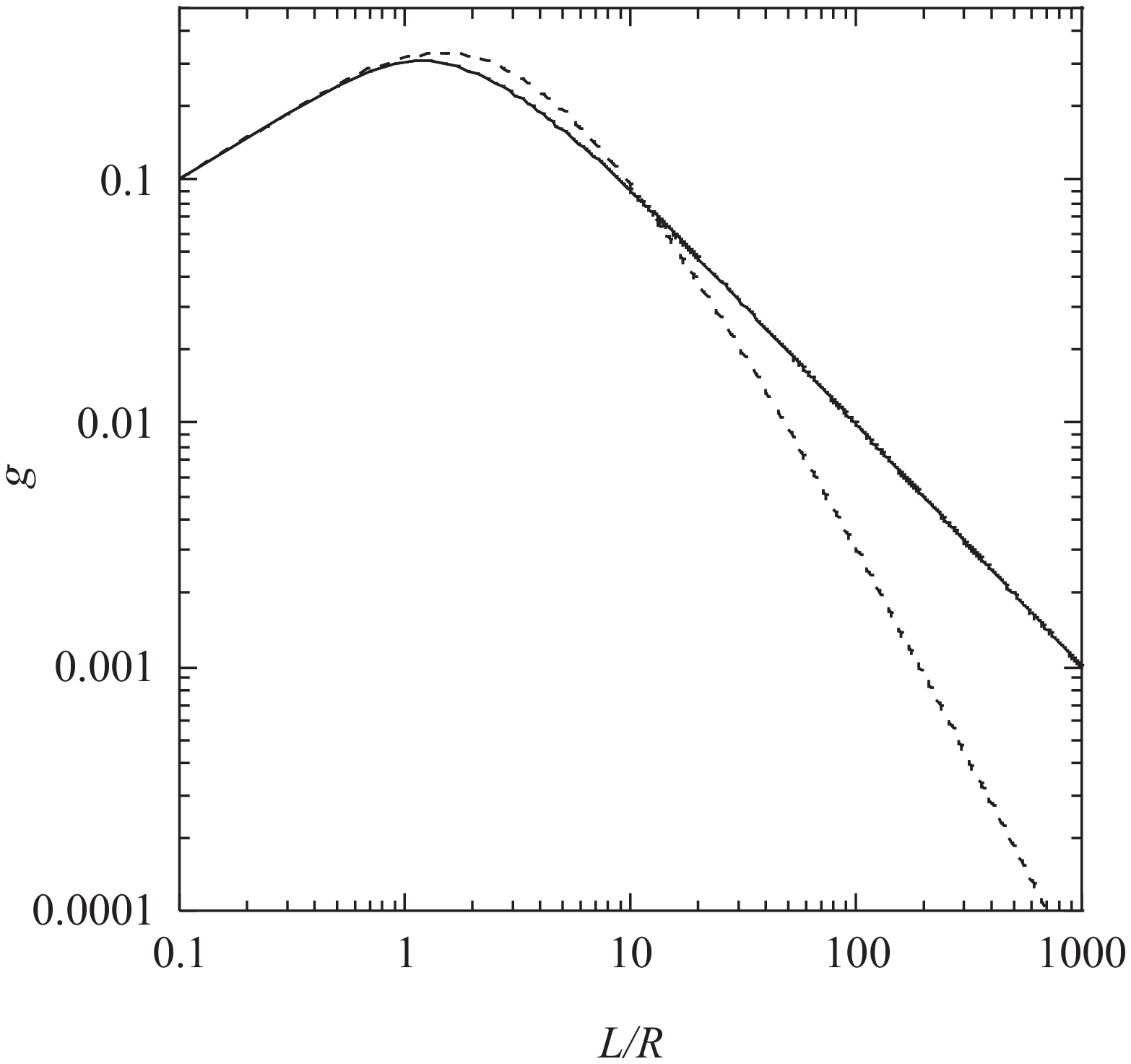}
\end{center}
\caption{} \label{Fi:4}
\end{figure}

\begin{figure}[tb]
\begin{center}
\includegraphics[scale=1]{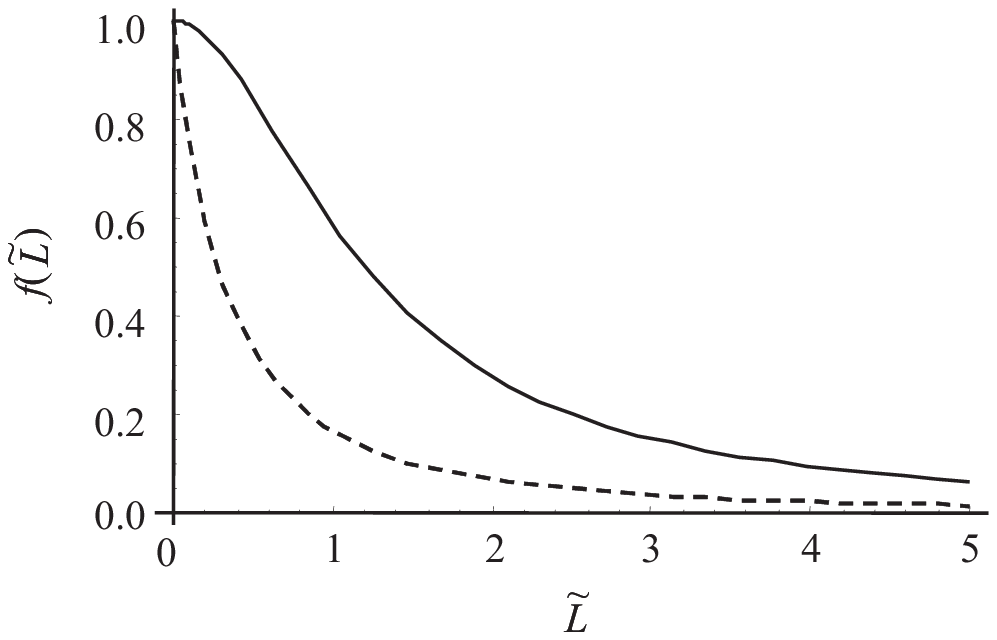}
\end{center}
\caption{} \label{Fi:5}
\end{figure}

\begin{figure}[tb]
\begin{center}
\includegraphics[scale=1]{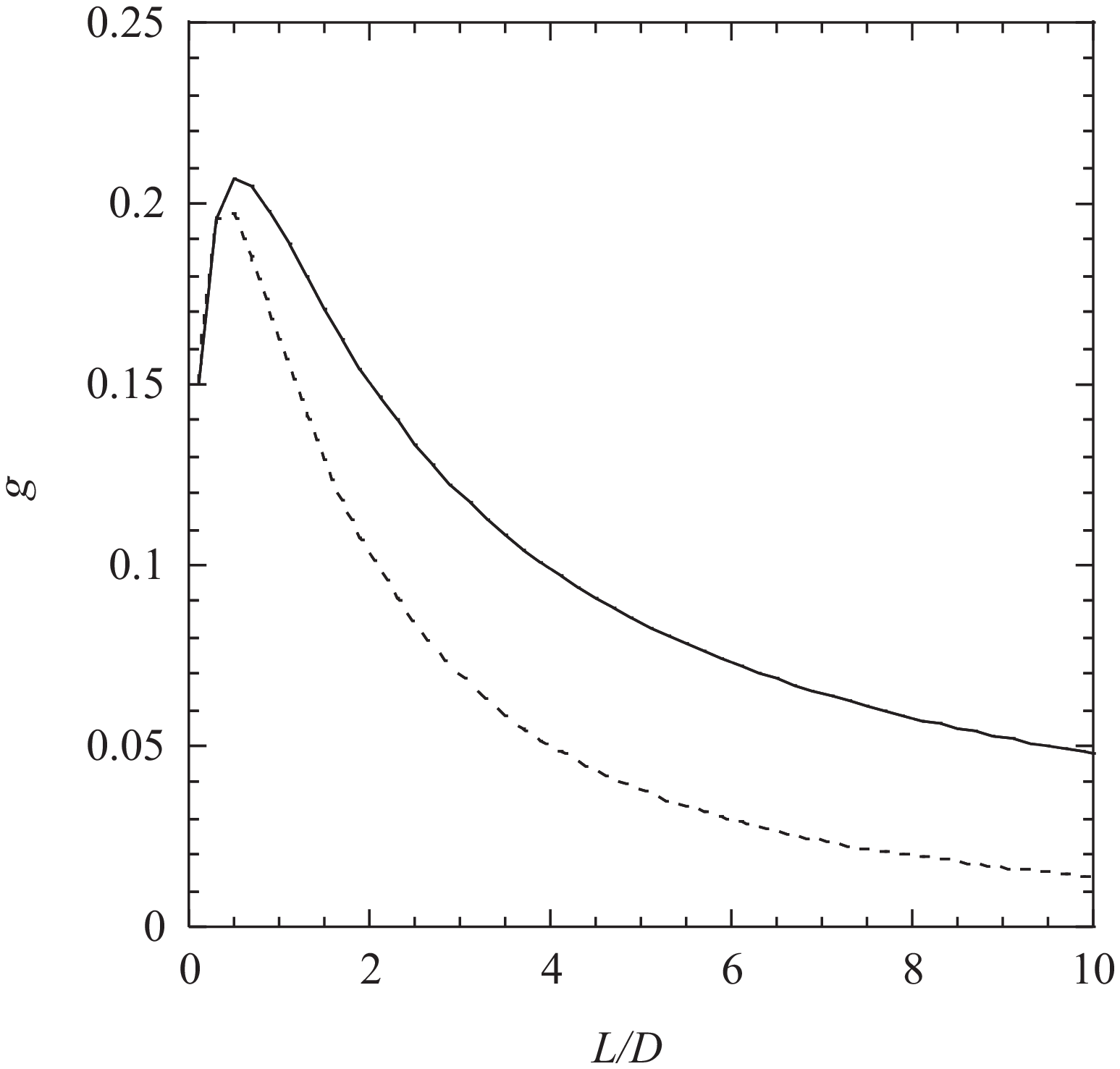}
\end{center}
\caption{} \label{Fi:6}
\end{figure}

\begin{figure}[tb]
\begin{center}
\includegraphics[scale=1]{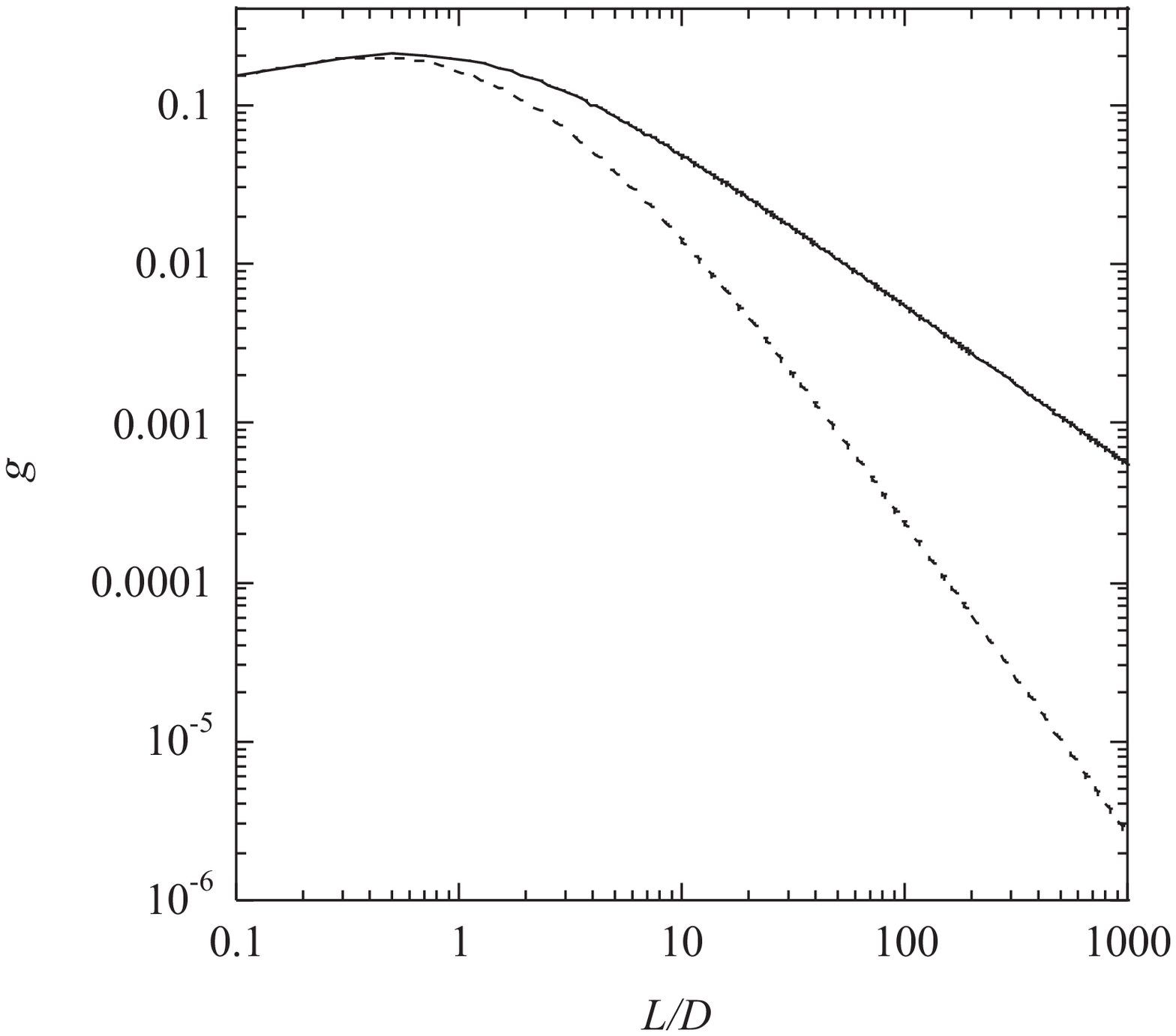}
\end{center}
\caption{} \label{Fi:7}
\end{figure}

\begin{figure}[tb]
\begin{center}
\includegraphics[scale=1]{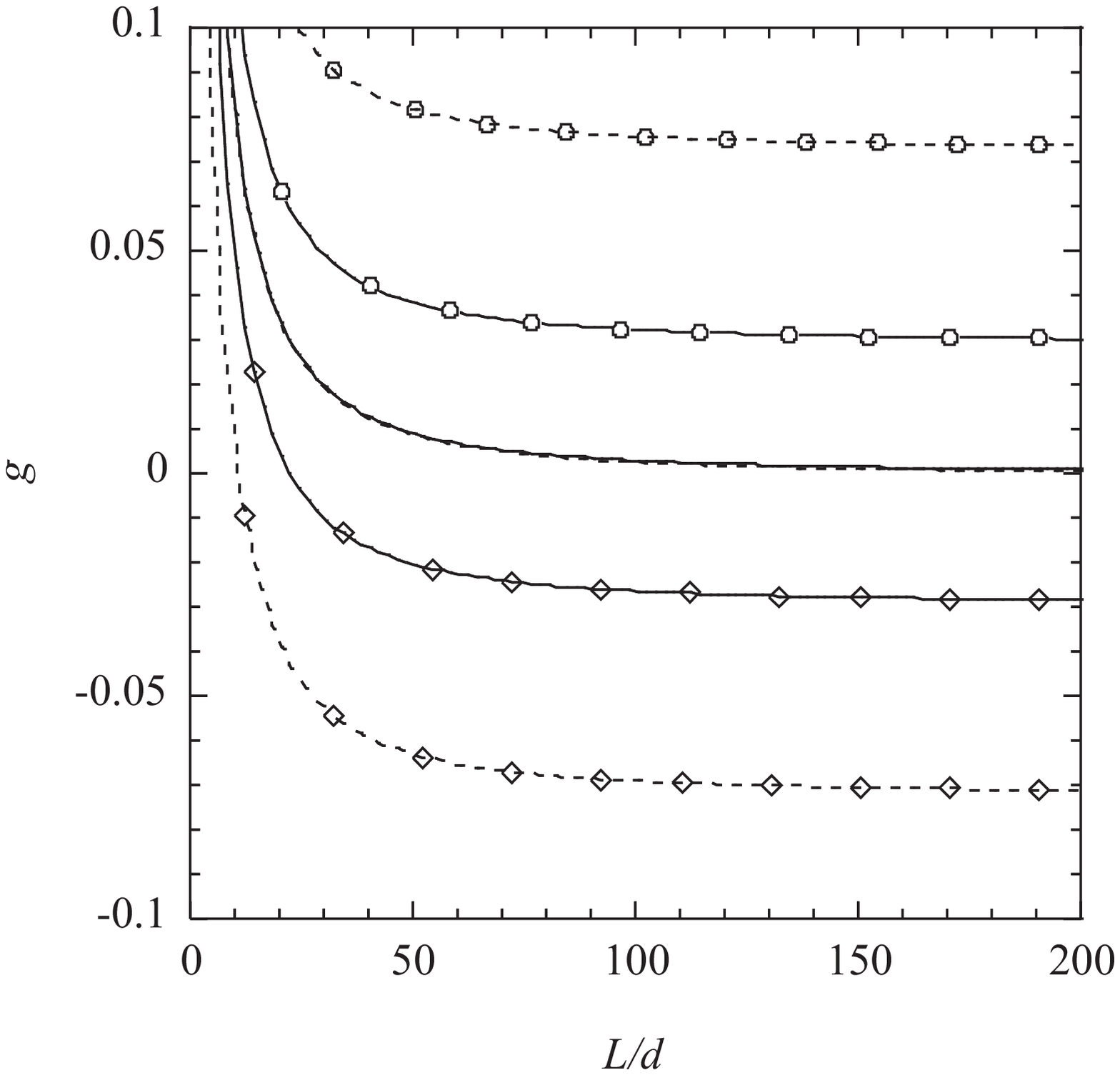}
\end{center}
\caption{} \label{Fi:8}
\end{figure}

FIG 1: The geometry of two conjugated polymers. $R$ is the
transverse separation and $D_0+D$ is the longitudinal shift. Also
shown are the sign of the transition densities for the lowest
excited exciton, the site labels `1' and `2' in a unit cell, and the
definitions of $a$, $d$, $L$, and $\theta$. For parallel chains
$D_0=0$, while for collinear chains $R=0$ and $D_0 = L$.
\\
\\
FIG 2: The dimensionless exciton transfer integral, $g(R,L) = J R^2
d/2 \mu_1^2$, evaluated for parallel chains ($D_0=D=0$) using the
line-dipole approximation with the plane-wave approximation Eq.\
(\ref{Eq:31}) (solid curve) and the standing wave approximation Eq.\
(\ref{Eq:29}) (dashed curve) for the exciton center-of-mass
wavefunction. The inter-chain separation $R =5d$, where $d$ is the
monomer length.
\\
\\
FIG 3: As for Fig.\ \ref{Fi:3}
\\
\\
FIG 4: The correction function, $f(\tilde{L})$, for the point-dipole
approximation. Solid curve: parallel chains where $\tilde{L} = L/R$;
dashed curve: collinear chains where $\tilde{L} = L/D$.
\\
\\
FIG 5: The dimensionless exciton transfer integral, $g(D,L) = JD^2
d/2 \mu_1^2$, evaluated for collinear chains ($R=0$ and $D_0=L$)
using the line-dipole approximation with the plane-wave
approximation Eq.\ (\ref{Eq:31}) (solid curve) and the standing wave
approximation Eq.\ (\ref{Eq:29}) (dashed curve) for the exciton
center-of-mass wavefunction. The chain separation $D =5d$, where $d$
is the monomer length.
\\
\\
FIG 6: As for Fig.\ \ref{Fi:6}.
\\
\\
FIG 7: The dimensionless exciton transfer integral, $g(R,L) = JR^2
d/2 \mu_1^2$, evaluated for parallel chains ($D_0=0$) using
transition densities (solid lines) and the line-dipole approximation
(dashed lines). The inter-chain separation, $R = d$. The
longitudinal shift, $D$, is $D= 0$ (circles), $D=a/2$ (no symbols),
and $D = a$ (diamonds).

\pagebreak

\end{document}